# Actinide Decontamination by Microwave Atmospheric Pressure Plasma jet: A Systematic Study Supported by Optical Emission Spectroscopy

R. Kar[1*], A. Bute[1], N. Chand[1], D.S. Patil[2], Romesh Chandra[3], P. Jagasia[4], P.S.Dhami[4], S.Sinha[1]

[1]Laser & Plasma Surface Processing Section, Bhabha Atomic Research Centre, Mumbai-400085

[2]Metallurgical Engineering and Materials Science, IIT Bombay, Mumbai-400076

[3] Accelerators and Pulsed Power Division, Bhabha Atomic Research Centre, Mumbai-400085

[4]Fuel Reprocessing Division, Bhabha Atomic Research Centre, Mumbai-400085

[*]Corresponding author: rajibkar.ph@gmail.com , phone: 91-22-2559-5001

**Abstract:** A single electrode microwave based atmospheric pressure plasma jet (APPJ) had been developed, characterized and applied for decontamination of Pu based synthetic radioactive waste. Argon plasma with small amount of $CF_4$ and $O_2$ was used for this purpose. The device was initially characterized by optical emission spectroscopy (OES) to determine its operational regime and used on Ta, a known surrogate of Pu for testing its efficacy in etching. Parametric optimization studies had been conducted thereafter on solid radioactive wastes of Pu and it was seen that presence of oxygen in plasma plays a crucial role in efficient removal of contamination. A scaled up multi-electrode version of this device was later designed and employed inside the glove box for similar studies. It was seen that ~ 92% decontamination could be achieved under optimized condition with both the devices and the scaled up APPJ device reduced operation time by 50%.

# 1. Introduction:

High demand for non-conventional energy sources for supply of uninterrupted clean energy has resulted in recent surge in nuclear industry. It is undeniable that nuclear power generation has shown remarkable progress over the last decades [1, 2]. However, with its fast expansion comes the question of nuclear waste management. It has always been a problem to completely remove radioactive contaminants from various components before final disposal of the waste. Conventional methods of decontamination like wet-chemical, mechanical machining and high pressure gas phase decontamination process normally generate a lot of secondary waste and also pose a possibility of accidental exposure and contamination to personnel [3-5]. These facts have necessitated a search for an alternative to traditional waste management system.

In any nuclear energy programme, volume reduction of radioactive waste is of paramount importance. Many researchers are working towards development of innovative treatment techniques for this purpose [6]. Nuclear wastes can be solid, liquid and gaseous in state. They are further classified based on category and amount of radioactivity. Presently, solid wastes are either incinerated or hydraulically compressed to achieve volume reduction. Liquid wastes go through a series of treatment including filtration, chemical treatment, ion-exchange, steam evaporation, solar evaporation and membrane processes (IAEA, 1983a). Gaseous radioactive wastes are trapped by different mechanisms like wet scrubbers, venturi, chillers and HEPA filters [6].

Since early nineties few research groups have reported plasma etching as an alternate option for decontamination of solid nuclear wastes [7-15]. Martz et al. [7] initiated this research by removing Pu using low pressure RF glow discharge. It was shown that plasma works at least 200 times faster than the conventional chemical processes. Other researchers like Kim et al. [12] used similar plasma and showed $CF_4$-$O_2$ plasma could convert $UO_2$ into $UF_6$ in two steps making handling of radioactive substance easier.

Among other related research, researchers of Hicks' group **[13, 14]** showed that atmospheric pressure 13.56 MHz $CF_4$–$O_2$ plasma could successfully etch Ta, a surrogate for Pu **[13]**. Later, Yang et al. **[8]** used similar plasma to etch $UO_2$ from a stainless steel surface using $CF_4$-$O_2$-He plasma. They concluded that atmospheric pressure plasma could be a promising technology for decontamination of transuranic wastes.

Windarto et al. **[5]** for the first time showed that microwave based atmospheric pressure plasma jet (APPJ) could be used for radioactive decontamination. They used $CF_4$-$O_2$ based microwave discharge for decontamination of radioactive $CoO_2$ from stainless steel surface at a power of 1.5 kWatts. Xie et al. **[15]** later conducted experiments with microwave APPJ at 1000 Watts and found that an optimum $O_2$ flow was required to increase etching effectiveness of the device.

These reports suggested that microwave based APPJ can serve as a potential alternative technique for radioactive decontamination on account of high chemical conversion rate, favorable economics, ease of operation, small size and low operational costs **[5, 15].**

In spite of these advantages, there exists an absolute scarcity of recent reports on plasma based nuclear waste management and this is where our research can contribute. In this study we have developed & characterized a prototype APPJ device and installed it inside a glove box for solid radioactive waste removal. Parametric optimization studies have been conducted on synthetic solid radioactive wastes. Based on its success a scaled up multi-electrode version was designed and employed inside the glove box. Both of these devices under optimized condition could successfully eliminate radioactive wastes and the scaled up APPJ device reduced the operation time by 50%.

## 2. Experimental:

Figure 1 shows a schematic describing the different phases in which this study was planned. Phase I of this study was focused on designing a microwave APPJ device. Phase II was divided in three different sets. In set I, important plasma parameters were measured with the help of optical emission spectroscopy (OES) and thermocouple. In set II, this device was subjected to extensive tests with different gas compositions. In set III, we used the device to etch Ta, a known Pu surrogate [14, 16, 17]. In phase III of our study, the device was installed inside a glove box for radioactive decontamination and parametric optimization study. After these experiments, we scaled up our device to a multi-electrode model in phase IV to address large area decontamination and used it inside glove box.

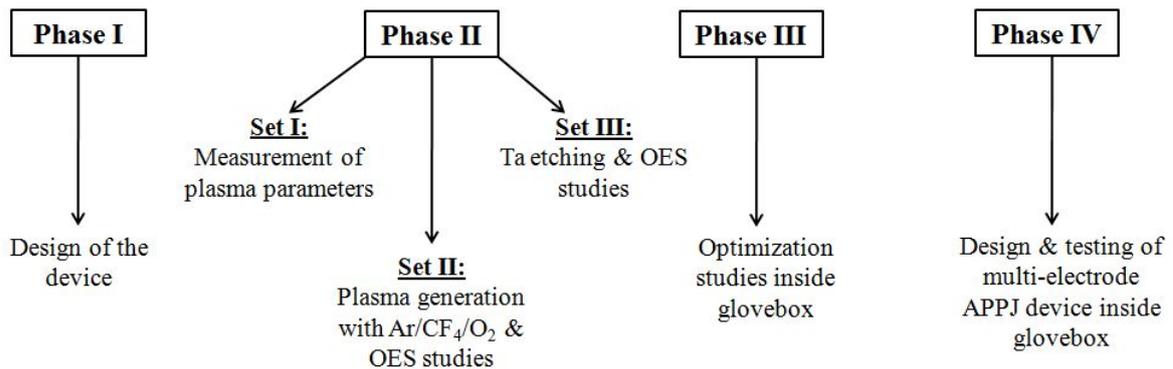

**Figure 1:** Schematic of the various experimental phases and their subsets.

Figure 2(a) shows the schematic of the set-up and actual set-up is shown in the Figure 2(b). A 2.45 GHz magnetron was used to generate microwave which then propagated through a waveguide. The APPJ device was connected to the waveguide through a LMR 400 co-axial cable for easy manoeuvering. Requisite gases for this experiment were passed through mass flow controllers and gas mixer before entering the device for plasma generation. Different components are duly labeled in the actual photograph.

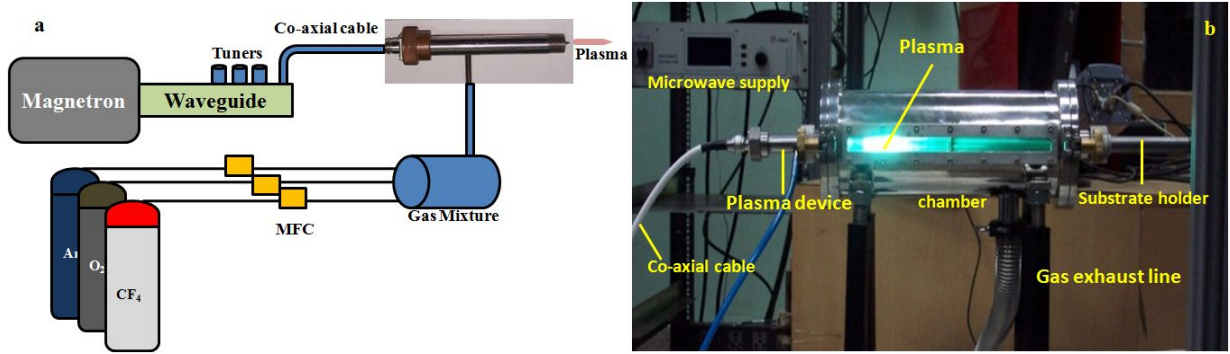

**Figure 2:** (a) Experimental schematic of the operational set-up of the microwave APPJ device, (b) photograph of the actual experimental set up.

**2.1 Phase I:** The design of the microwave APPJ device was based on co-axial structure **[18]**. Low production cost, ease of replication with minimum machining skills were some of the features which were considered for designing this device. The most important factor during design was to match the characteristic impedance of the device with that of the co-axial cable (50 Ohm). Equation 1 was used to calculate the impedance of the device [**18**].

$$Z_0 \approx (138/\sqrt{\varepsilon_r}) \log_{10}(D/d) \qquad (1)$$

Where $\varepsilon_r$ is the relative permittivity, D is the inner diameter of the outer tube and d is the diameter of the inner rod. In ambience, $\varepsilon_r$ is taken as 1. Values of D and d during the design were fixed as 140 and 6.35 mm respectively which made the characteristic impedance of the device ~ 47.4 Ohm.

Photograph of the APPJ device is shown in Figure 3. Figure 3(a) shows different component parts of the device while Figure 3(b) shows the device in assembled form. N-type connector at the device end shown in the figure 3(a) was connected to the LMR 400 coaxial cable.

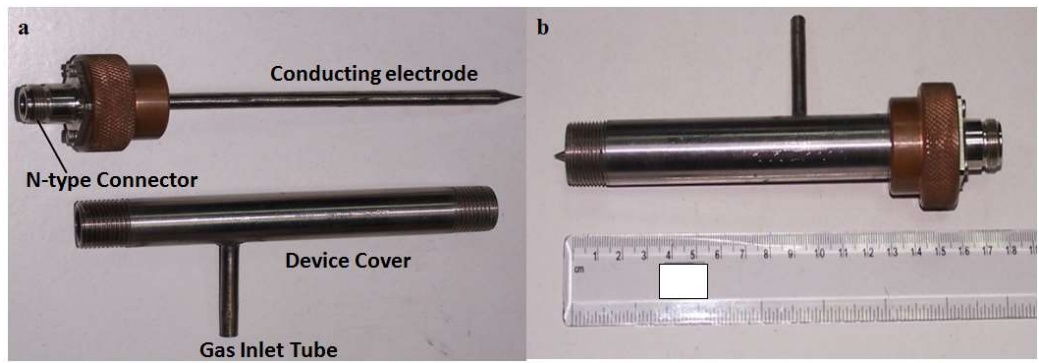

**Figure 3**: (a) Different parts of the APPJ device, (b) The assembled APPJ device.

**2.2 Phase II:** For the experiment in set I, a mixture of 3.5 LPM Ar and 100 SCCM $H_2$ was used to generate the plasma. Electron temperature ($T_e$) and number density ($n_e$) of the plasma were measured with the help of OES. Details of this measurement procedure are given in section 3.1.

For set II a gas mixture of Ar and $CF_4$ was chosen, and $O_2$ was added in some cases. All these experiments were conducted in a closed stainless steel (SS) 316 chamber as decomposition of $CF_4$ could release F/ $F_2$ in plasma. Exhaust from the chamber was immersed in a KOH filled container to convert escaped $F_2$ into KF. Introduction of $CF_4$ gave rise to a distinctive green flame in plasma as seen the Figure 2(b). OES studies were done to monitor the changes occurring within plasma due to changes in gas composition.

In set III, two different experiments were conducted on Ta surrogate both in presence and in absence of $O_2$ in plasma. Weights of the Ta substrates were measured both before and after the plasma etching experiments to compare etching rates.

**2.3 Phase III:** After successful plasma generation in laboratory, the device was transferred into a glove box for radioactive decontamination experiments and parametric optimization studies. All experiments were conducted with net microwave power ranging between 80-100 watts. Pu was chosen as the radioactive source since it has one of the highest specific activities among all transuranic elements. Several synthetic samples of Pu were prepared on different SS 304L discs of ~25 mm dia. One ZnS(Ag) scintillation counter having efficiency ~ 25% (standardized against $^{241}$Am standard source) was used for measuring α counts before and after each experiment. Decontamination factor (DF) was calculated using equation 2. Higher DF value signifies better removal of waste.

DF (%) = (Initial CPM – Final CPM) / Initial CPM x 100        (2)

**2.4 Phase IV:** Based on our results using a single electrode APPJ device, a multi-electrode APPJ device was designed in phase IV to scale up the waste removal process. Design of this device involved prior simulation as analytical solutions were unavailable for impedance matching for such a device. The design and experimental results obtained with this device will be discussed section 3.5.

**3. Results:**

**3.1 Phase II, set I:** As stated in previous section, OES was used to measure electron temperature ($T_e$) and number density ($n_e$) of plasma. Partial local thermodynamic equilibrium (PLTE) model was assumed during these measurements since plasma was operating at atmospheric pressure **[19, 20]**. Equation 3 based on two-line ratio method of Boltzmann model was used for estimation of $T_e$.

$KT_e = E_{2*2} - E_{1*1} / \ln[(I_{1*1} \lambda_{1*1} A_{2*2} G_{2*2}) / (I_{2*2} \lambda_{2*2} A_{1*1} G_{1*1})]$        (3)

Schematic shown in Figure 4 depicts the line transition between two excited levels ($2^*/1^*$) and their downward transition levels (2/1).

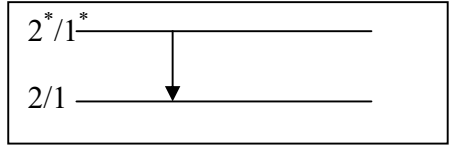

**Figure 4**: Schematic showing atomic transitions between two different levels.

In equation 3, K is the Boltzmann constant, $T_e$ is the electron temperature, $E_{2*2}$ & $E_{1*1}$ are energies of the emitted radiations from the excited levels, $I_{2*2}$ & $I_{1*1}$ are the integrated intensities of the emitted radiation. $\lambda_{2*2}$ & $\lambda_{1*1}$ are wavelengths of the emitted radiation. A & G parameters signify Einstein's coefficient and degeneracy respectively. NIST database has been used to find out the values of E, A and G for different levels of Ar.

$n_e$ was calculated from the Stark broadening of $H_\beta$ spectral line which follows a Lorentzian line profile. For obtaining $n_e$, this line is fitted with a Voigt profile and the FWHM contribution from the Lorentzian part is taken for calculations. Following work of Gigosos et al. **[21]** $n_e$ can be estimated from equation 4 given below.

$$H_{\beta\ FWHM}\ (nm) \approx 4.8 \times (n_e/10^{-23}\ m^{-3})^{0.681} \qquad (4)$$

Figure 5(a) shows OES spectrum used for $T_e$ and $n_e$ measurements. Ar lines are clearly visible along with prominent $H_\alpha$ line at 656 nm. Fitting of $H_\beta$ line for calculation of $n_e$ is shown in Figure 5(b) where actual spectrum is shown as black while the fitted one shown in red. The inset box provides all information about the fitting parameters.

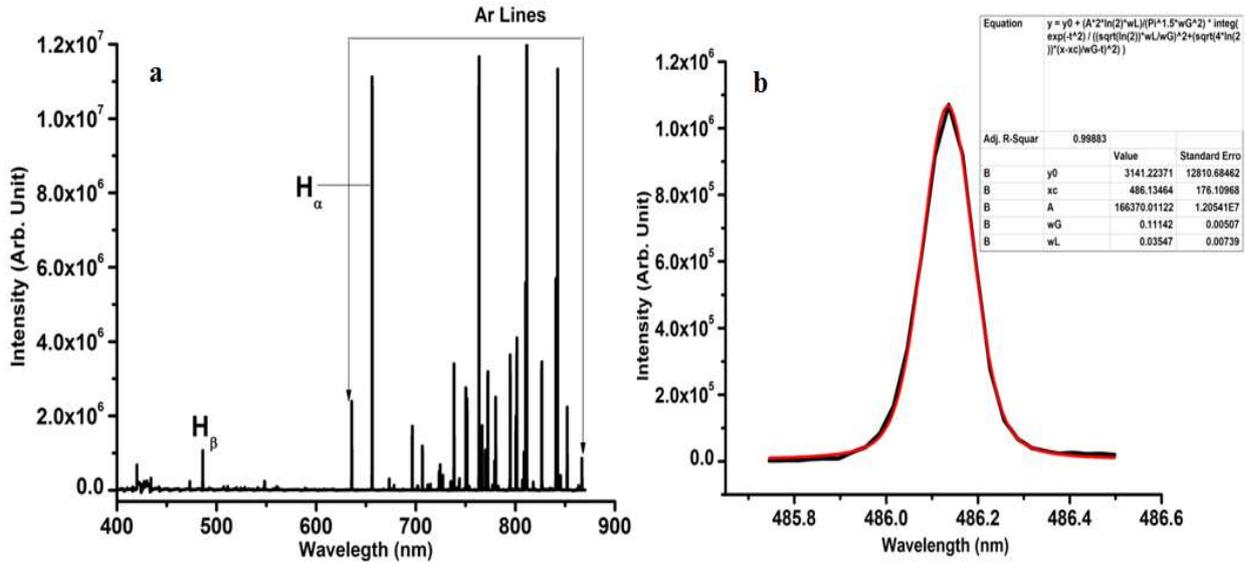

**Figure 5:** (a) The obtained emission spectrum for electron temperature and number density measurement, (b) detailed profile of $H_\beta$ line with Voigt fitting.

One K-type thermocouple was used to measure the flame temperature of plasma at the tip and at centre of the plume. Table 1 shows all measured plasma parameters averaged over three different measurements.

**Table 1:** Different plasma parameters measured with thermocouple and emission spectroscopy.

| Plasma parameters | Value |
|---|---|
| Electron temperature | 5334°C |
| Number density | $7.4 \times 10^{-19}$ m$^{-3}$ |
| Maximum flame tip temperature | 62°C |
| Maximum plasma core temperature | 325°C |

**3.2 Phase II, set II:** For OES measurements, $O_2$ gas was systematically varied in Ar/ $CF_4$ plasma and simultaneous changes in plasma was studied. Figure 6(a) shows OES spectrum of plasma produced with Ar + 0.5% $CF_4$. The main feature of the spectrum is prominent presence of CN violet system at 388.3, 420 and 358 nm [**22**]. $C_2$ Swan system is also observed near 516 nm, ~ 474 and 563.5 nm [**23**]. Ar lines appeared 700 nm onwards in this spectrum. Figure 6(b) shows the OES spectrum where $CF_4$ and $O_2$ are mixed in 4:1 ratio. Figure 6(c) depicts OES spectrum with $O_2$ flow increased while maintaining $CF_4$: $O_2$ at 1:1 ratio. Figure 7 depicts the truncated OES spectrum of Figure 6(c) with wavelength restricted between 350-600 nm. The spectrum clearly shows presence of CO. Major CO bands from Angstrom system (~ 561 nm, 520 nm, 483 nm and 451 nm) and flame system (~ 448 nm, 458 nm, 493 nm, 498 nm, 503 nm, 532 nm and 544 nm) are identified [**22**]. Major $CO^+$ band of Baldet-Johnson system is also observed at ~ 395 nm. Bands of $CO_2$ from Fox, Duffendack and Barker's system were seen between 350-365 nm. Most intense molecular bands of $N_2$ first positive system is seen between 575-580 nm while most intense band of $N_2^+$ was also seen ~ 423-427 nm [**22**]. Strong emission lines of C I (~ 427 nm and 437 nm) [**23**] and O I could also be identified in this spectrum.

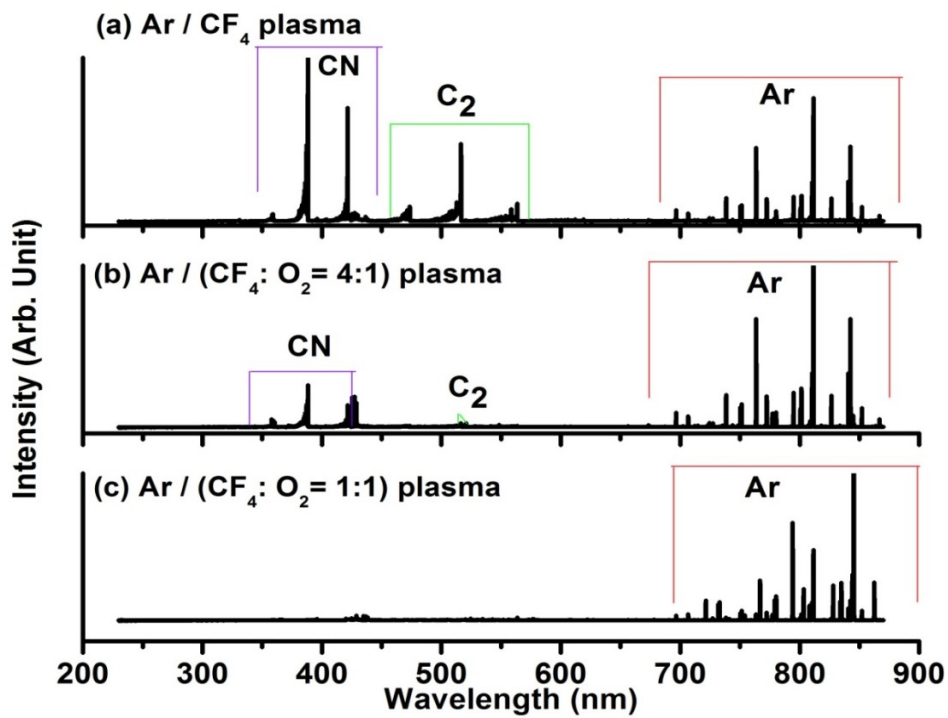

**Figure 6:** OES spectrum of (a) Ar / $CF_4$ plasma, (b) Ar / ($CF_4$: $O_2$ = 4:1) plasma, (c) Ar / ($CF_4$: $O_2$ = 1:1) plasma.

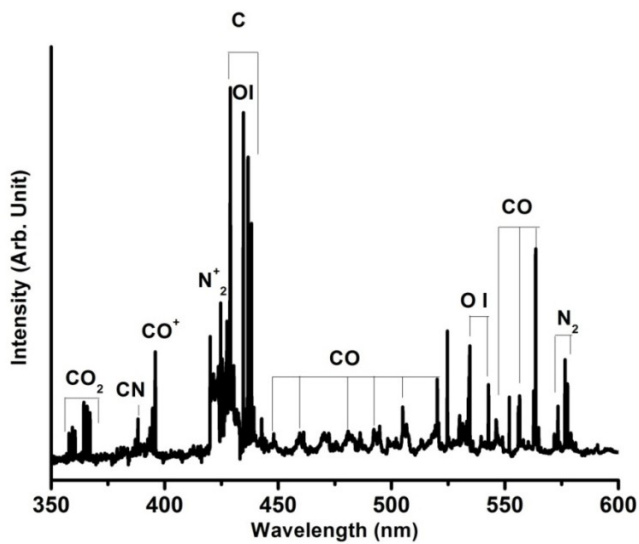

**Figure 7:** Truncated OES spectrum from figure 6(c) between 350-600 nm.

Due to the absence of detectable signatures of F in OES spectra, concentration of $CF_4$ in plasma was further increased from 0.5% to 1.5% of Ar. Thereafter, another experiment was conducted by mixing $O_2$ in the plasma where $CF_4:O_2$ was kept in 4:1 ratio. In both cases major F I peaks were detected at ~ 686 nm, 704 nm and 713 nm. Figure 8 shows OES spectrum showing presence of these F I lines.

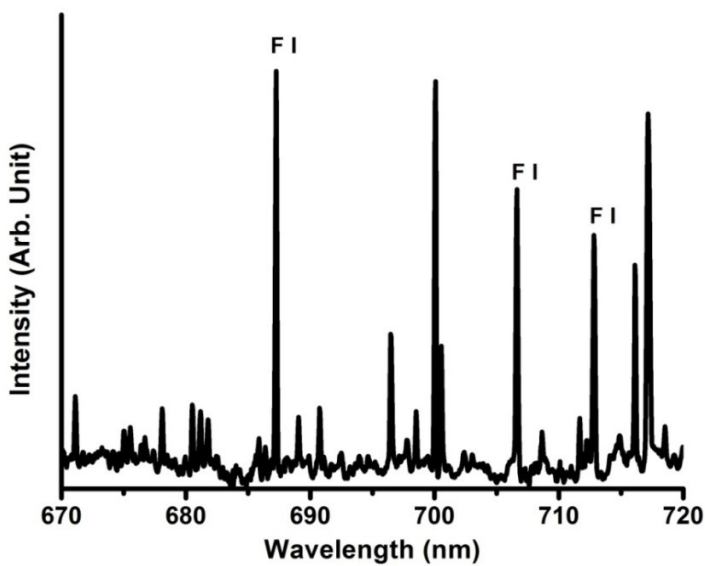

**Figure 8:** OES spectrum from APPJ showing signature of F I emission lines.

**3.3 Phase II, set III:** During Ta etching experiments, $CF_4$ concentration was further increased to 3% of Ar flow and two sets of experiments were conducted for 45 minutes each. The first set had a gas mixture of 3% $CF_4$ + Ar while $O_2$ was added in the second set to the above mixture to study its effect on etching [**15, 24**].

Figure 9(a) shows the SEM image of Ta surface before etching while Figures 9(b) & 9(c) shows the surface after etching without and with $O_2$ respectively. Comparing Figures 9(b) & 9(c), it is clearly seen that $O_2$ makes plasma etching more aggressive. In oxygen less condition, the weight of Ta substrate reduced from 1.75 gm to 1.70 gm indicating only 50 mg loss of weight. With introduction of oxygen, weight of Ta went down from 1.69 gm to 1.43 gm indicating weight loss of 256 mg indicating an enhanced etching rate of 4.27 mg/ min.

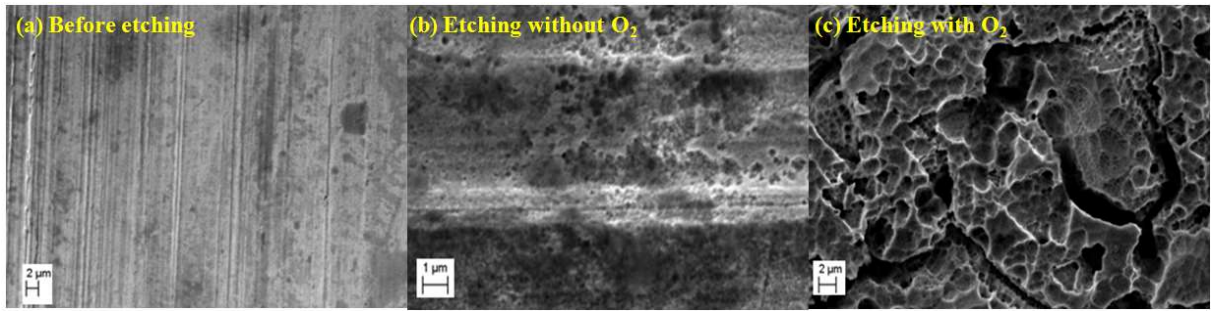

**Figure 9:** SEM images of (a) Ta surface before etching, (b) Ta surface after etching in Ar+CF$_4$ plasma, (c) Ta surface after etching in Ar+CF$_4$+O$_2$ plasma.

Figure 10 (a, b) shows the OES spectra recorded over two spectral regions during the etching of Ta substrate. The spectrum in Figure 10(a) shows clear presence of CF$_2$ emission bands between 240-270 nm regions in both the cases, without and with oxygen. Presence of CF$_2$ bands signifies dissociation of CF$_4$ and release of Fluorine for etching of Ta substrates. Figure 10(b) shows another section of spectrum between 690-800nm. Differences between the two spectra shown in Figure 10(b) are marked by blue rectangles in the figure. The first difference is the emergence of O I peak ~ 777.4 nm in the bottom graph of 10(b) due to introduction of oxygen. Other difference is presence of a new F I peak at ~ 690 nm in the bottom graph suggesting enhanced dissociation of CF$_4$ in presence of O$_2$. Prominent F I peaks seen in both the spectra occur at ~ 702, ~ 713 and ~ 780 nm.

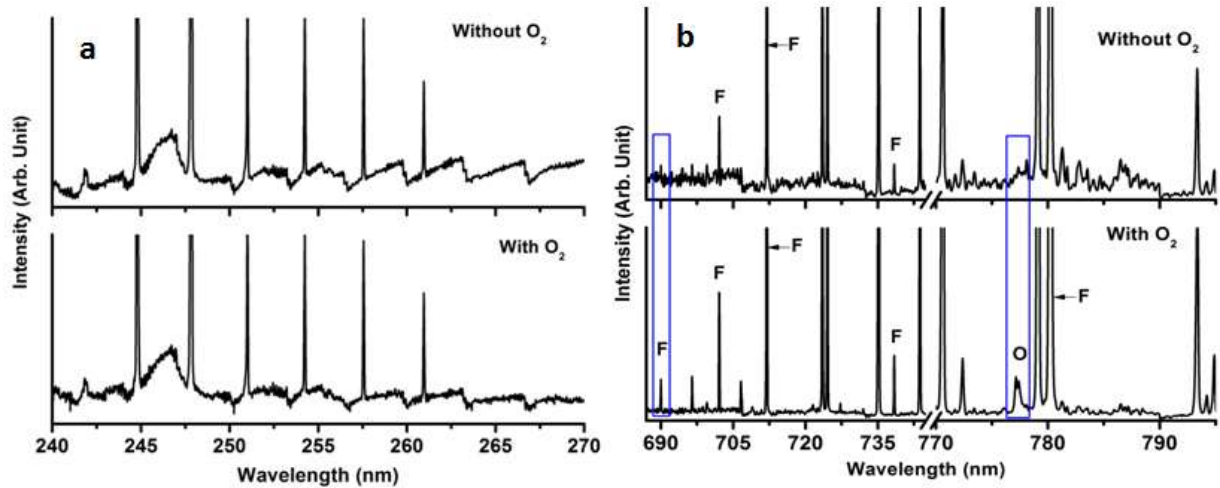

**Figure 10:** OES spectra taken over two spectral regions during Ta etching experiment showing presence of (a) $CF_2$ emission between 240-270 nm, (b) emission of F, O and Ar peaks between 690-800 nm.

**3.4 Phase III:** After successful etching of Ta surrogates, the APPJ device was installed inside a radioactive glove box for carrying out decontamination experiments on active samples. Synthetic radioactive samples of Pu having α activity of 40,000 CPM were prepared by taking 20 μl $Pu(NO_3)_4$ drops on the centre of SS 304L discs (25 mm diameter). These samples were then exposed in plasma to different gas compositions for 10 minutes to study the decontamination efficiency.

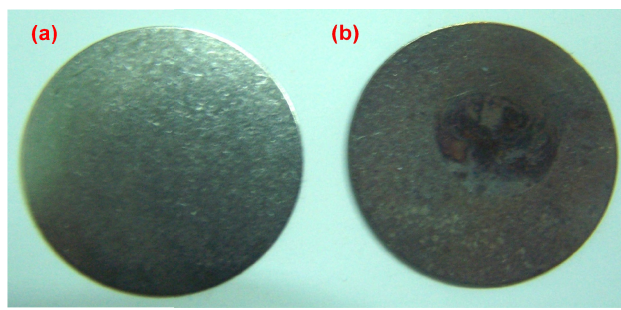

**Figure 11:** (a) Pristine SS304L disc, (b) Synthetic radioactive $Pu(NO_3)_4$ sample

To systematically evaluate effect of oxygen on decontamination of α contaminated samples, we conducted several experiments inside glove box. A set of experiments were conducted with $CF_4$ concentration increased from 1% to 4% in absence and in presence of only 0.1% oxygen while DF was evaluated after each experiment. The result is graphically displayed in Figure 12(a) where it is seen that DF increases considerably in presence of oxygen with the maximum increase being 32.5% for 4% flow of $CF_4$. It is also seen that with increasing $CF_4$, gap between "Without $O_2$" and "With 0.1% $O_2$" values widens as $CF_4$ increases. Since 0.1% oxygen brought remarkable changes in DF values we decided to investigate the effect of oxygen concentration in further detail. For this purpose $CF_4$ was set as 3% of total gas flow as maximum DF was observed at this value (Figure 12(a)). From Figure 12(b) it is seen that when oxygen was increased from 1% to 5% of the total gas mixture, there is a linear increase in DF and no saturation in its value is observed in the investigated experimental regime. It is seen that DF reaches 90% and beyond when oxygen flow rate was kept between 3.5% to 5%. However, higher percentage of oxygen flow created instabilities in plasma jet. Hence, oxygen flow was restricted within 3% to 4% of total gas flow for stable operation of APPJ device for efficient decontamination.

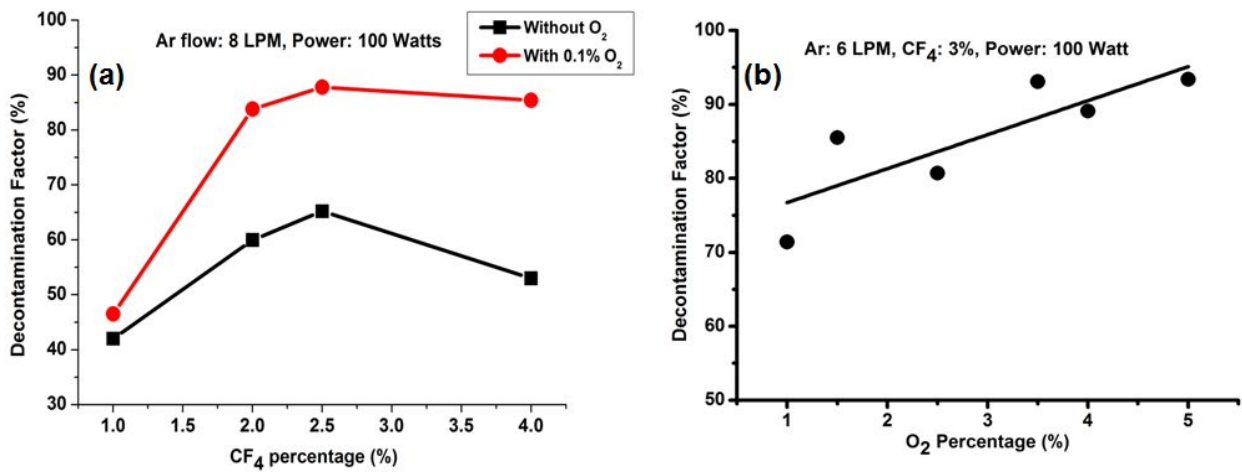

**Figure 12:** Effect of $O_2$ (a) on the decontamination factor, (b) flow rate variation on the decontamination factor.

Argon being the plasma medium, it is expected to play a major role in decomposition of $CF_4$ and $O_2$ facilitating the etching process. So, to investigate the effect of argon flow rate systematic experiments were conducted varying Ar flow rate from 12 LPM to 3 LPM. Below 3 LPM plasma could not be sustained in the AAPJ device. Figure 13 shows that with decreasing Ar flow rate DF continuously increases and below a flow rate of 5 LPM DF value almost saturates reaching as high as 98%. However, below 6 LPM Ar flow rate, plasma jet showed continuous fluctuation & increased instability.

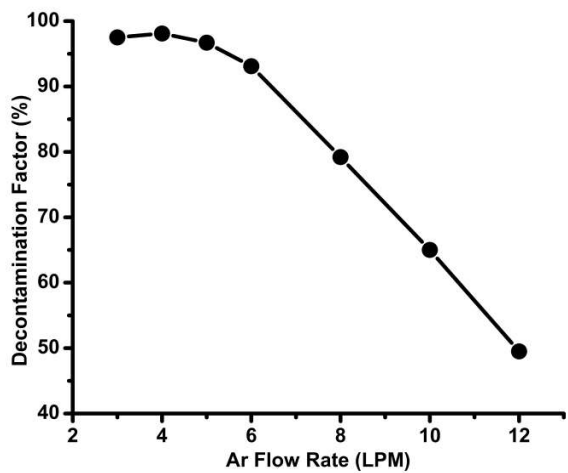

**Figure 13:** Effect of Ar flow rate variation on the decontamination factor.

**3.5 Phase IV:** Based on the encouraging results obtained with the single electrode APPJ device, a multi-electrode model of the device was conceptualized for scaling up of the decontamination process. However, designing a multi-electrode AAPJ device by analytical means, with appropriate matching of impedance not being simple as single electrode device (section 2.1) the device was simulated using time domain and frequency domain solver simulations in CST

Microwave Studio. During simulation, a three-electrode APPJ device was designed with proper impedance matching. In the simulation geometry a waveguide port was defined at the input of N-type connector for supplying 2.45 GHz microwave to the device with power transmitted being absorbed completely by another waveguide port located at the other end. Figure 14(a) shows impedance variation of the device along the length of the simulated device while Figure 14(b) shows the fabricated device operational in laboratory.

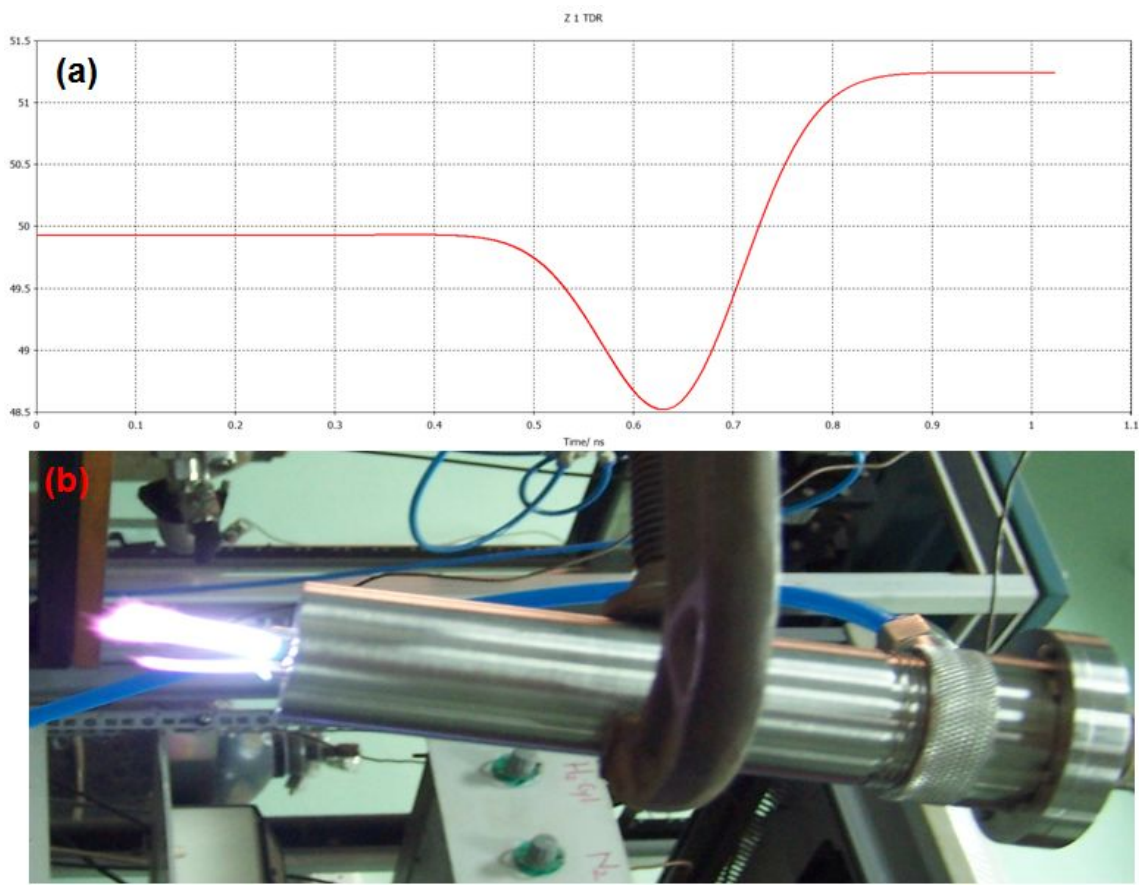

**Figure 14:** (a) Dependence of impedance along the length of the device, (b) fabricated operational device in lab.

After initial testing, the device was transferred inside glove box to test its effectiveness in nuclear waste decontamination. Similar synthetic α waste as mentioned in section 3.4 were prepared for these studies. Experiments were performed for 5 minutes each keeping $CF_4:O_2$ at 1:1 ratio and Ar flow rate was varied systematically from 18 to 6 LPM. Below 6 LPM as in the case of a single electrode APPJ, plasma jet in multi-electrode device too became unstable. Figure 16 shows the variation of Ar flow rate with decontamination factor. From the figure it is seen that DF increases linearly with decreasing Ar flow rate. Our results showed that almost 95% decontamination can be achieved within 5 minutes at 8 LPM Ar flow. When compared with the performance of a single electrode device, it is seen that the scaled up multi-electrode model can reduce operation time by almost 50%.

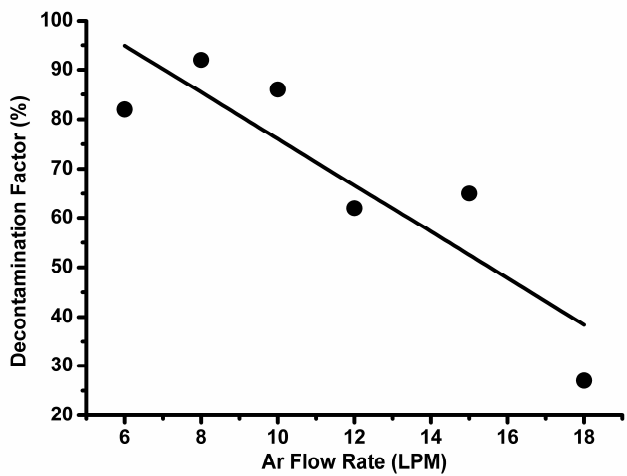

**Figure 16:** Variation of Ar flow rate variation on the decontamination factor for multi-electrode APPJ device.

**4. Discussion:** In the work being reported here, OES has been used extensively to narrow down operational regime of a plasma device before deploying the device inside a glove box where actual parametric optimization has been carried out. From Figures 6 and 7, it is seen that introduction of oxygen in plasma resulted in suppression of CN and $C_2$ bands in the spectra. In absence of oxygen, dissociated carbon from $CF_4$ formed CN in plasma after reacting with atmospheric nitrogen whereas, in presence of oxygen, this carbon reacted rapidly with oxygen to form CO, $CO_2$ and CO+. Unlike some previous reports no signatures of $CF_2$ / $CF_3$ emission or Fluorine was observed in these spectra [25, 26]. Absence of $CF_2$ or $CF_3$ might be due to the toughened glass window used in our experimental set up which could have blocked these emission bands between 150-250 nm. Absence of F in the spectra may be attributed to the fact that, at atmospheric pressure excessive collisions of atomic F with heavy particles could have quenched spontaneous radiation of atomic F [27]. Hence signatures belonging to these species might not be identifiable via OES here unlike in low pressure plasmas [25, 26]. Realizing that detectable signature of F could be correlated to efficient etching we increased the $CF_4$ content in the gas mixture from 0.5% to 1.5% till major F peaks in OES spectrum (Figure 8) were detected. During decontamination experiments inside glove box, for a given plasma power, efficient decomposition of $CF_4$ occurred with the introduction of oxygen. This observation correlates well with our OES result (Figure 10(b)) where additional F peak was observed in presence of oxygen suggesting enhanced dissociation of $CF_4$ in presence of $O_2$ in plasma. Therefore, our observations appear to be in agreement with observations recorded by Setareh et al. [24].

Based on extensive experiments conducted with APPJ device and considering factors like increase in DF, plasma stability and gas economy, we have arrived at optimized operational parameters corresponding to 6 LPM Ar flow, with $CF_4$ & $O_2$, each kept at 3% to 3.5% of the total gas flow, preferably at 1:1 ratio with input power of 100 Watts. Removal of hazardous nuclear waste from solid surfaces of SS upto 92% was observed within 10 minutes for a single electrode APPJ device and 5 minutes with a multi-electrode device under optimized conditions. The material balance was also found to be better than 98% in all the cases.

To validate our measured data on alpha activity post plasma etching results, a few samples after decontamination were dissolved completely in 8 M HCl and the resulting solution of known volume was subjected to radiometry to measure α counts. It was seen that number of counts obtained after the decontamination experiments matched well with what was obtained in the acidic solution. This experiment confirmed the fact that fluorination actually removed contamination from the surface.

Plasma probably is the most underused tool applied towards nuclear waste management system. Limitations in scaling up often said as a reason for restricting widespread application of this technology [7, 9, 11]. Our results suggest possibility of scaling up such devices & making it possible to put plasma based nuclear waste management in practice.

**5. Conclusion:** We have designed and developed an APPJ device for solid nuclear waste removal. Plasma characteristic of this device has been analyzed by OES. Employing a device inside a glove box we have successfully achieved removal of hazardous radioactive contaminant. It is seen that presence of oxygen in plasma has a very crucial role in efficient etching of contaminated surfaces. Optimized Ar flow rate of 6 LPM ensured stable plasma jet and high decontamination efficiency. Pu was used as a source of α- activity and with the help of APPJ device upto 92% decontamination factor was achieved under optimized condition. Subsequently, with a scaled up multi-electrode APPJ operational time for decontamination could be reduced by 50%. Portability, ease of operation, high decontamination efficiency and very low secondary waste generation makes this device a potential alternative in solid nuclear waste management.

**References:**

[1] S. Taylor, The Fall and Rise of Nuclear Power in Britain: A History, UIT Cambridge, ISBN 1906860319, 9781906860318, 2016.

[2] W. Conard Holton, Power Surge: Renewed Interest in Nuclear Energy ,Env. Health Pers., Vol 113, Number 11, 2005, A743-A749.

[3] Y.H. Kim, Y.H. Choi, J.H. Kim, J. Park, W.T. Ju, K.H. Paek, Y.S. Hwang, Decontamination of radioactive metal surface by atmospheric pressure ejected plasma source, Surf. Coat. Tech. 171 (2003) 317–320.

[4] D.E. Roberts, T.S. Modise, Laser removal of loose uranium compound contamination from metal surfaces, Appl. Surf. Sci. 253 (2007) 5258-5267.

[5] H.F. Windarto, T. Matsumoto, H. Akatsuka, M. Suzuki, Decontamination Process Using CF4-O2 Microwave Discharge Plasma at Atmospheric Pressure, J. Nuc. Sci. Tech. 37 (2000) 787-792.

[6] K. Raj, K.K. Prasad, N.K. Bansal, Radioactive waste management practices in India, Nucl. Eng. Des. 236 (2006) 914–930.

[7] J.C. Martz, D.W. Hess, J.M. Haschke, J.W. Ward, B.F. Flamm, Demonstration of plutonium etching in a $CF_4/O_2$ RF glow discharge, J. Nuc. Mater. 182 (1991) 277-280.

[8] X. Yang, M. Moravej, S.E. Babayan, G.R. Nowling, R.F. Hicks, Etching of uranium oxide with a non-thermal, atmospheric pressure plasma, J. Nuc. Mat. 324 (2004) 134–139.

[9] R.F. Hicks, H.W. Herrmann, Atmospheric-Pressure Plasma Cleaning of Contaminated Surfaces, US dept. of energy, Project Number: 73835 Grant Number: FG07-00ER45857, 2003.

[10] J.M. Veilleux, M.S. El-Genk, E.P. Chamberlin, C. Munson, J. FitzPatrick, Etching of $UO_2$ in $NF_3$ RFplasma glow discharge, J Nucl. Mater. 277 (2000), 315

[11] J.M. Veilleux, Y. Kim, Can Plasma Decontamination Etching of Uranium and Plutonium be Extended to Spent Nuclear Fuel Processing, Los Alamos National lab 52nd Annual Meeting July 17-21, 2011.

[12] Y. Kim, J. Min, K. Bae, M. Yang, Uranium dioxide reaction in $CF_4/O_2$ RF plasma, J.Nuc. Mater. 270 (1999) 253-258.

[13] A. Schutze, J.Y. Jeong, S.E. Babayan, J. Park, G.S. Selwyn, R.F. Hicks, The Atmospheric-Pressure Plasma Jet: A Review and Comparison to Other Plasma Sources, IEEE Trans. Plasma Sci. 26 (1998) 1685-1694.

[14] V.J. Tu, J.Y. Jeong, A. Schutze, S.E. Babayan, G. Ding, G.S. Selwyn, R.F. Hicks, Tantalum etching with a nonthermal atmospheric-pressure plasma, J. Vac. Sci. Technol. A 18 (2000) 2799-2805.

[15] H.D. Xie, B. Sun, X.M. Zhu, Y.J. Liu, Influence of $O_2$ on the $CF_4$ Decomposition by Atmospheric Microwave Plasma, Int. J. Plasma Env. Sci. Tech. 3 (2009) 39-42.

[16] H. Zheng, F.Y. Yueh, T. Miller, J.P. Singh, K.E. Zeigler, J.C. Marra, Analysis of plutonium oxide surrogate residue using laser-induced breakdown spectroscopy, Spec. Act. Part B 63 (2008) 968–974.

[17] A.J. Duncan, Evaluation of Possible Surrogates for Validation of the Oxidation Furnace for the Plutonium Disposition Project, Savannah River National Laboratory, WSRC-TR-2007-00471, Rev. 0, January 2008.

[18] M. Jasinski, Z. Zakrzewski, J. Mizeraczyk, New atmospheric pressure microwave microplasma source, Acta Technica CSAV 53 (2008) 347-354.

[19] X. Zhu, W. Chen, Y. Pu, Gas temperature, electron density and electron temperature measurement in a microwave excited microplasma, J. Phys. D: Appl. Phys. 41 (2008) 105212 (6pp).

[20] Q. Zhang, G. Zhang, S. Wang, L. Wang, N. Huo, Characterization of an atmospheric-pressure helium plasma generated by 2.45-GHz microwave power IEEE transac. Plas. Sci., 38 (2010) 3197-3200.

[21] M.A. Gigosos, M.A. Gonzalez, V. Cardenoso, Computer simulated Balmer-alpha, -beta and -gamma Stark line profiles for non-equilibrium plasmas diagnostics, Spec. Act. Part B 58 (2003) 1489–1504.

[22] R.W.B. Pearse, A.G. Gaydon, The identification of molecular spectra, John Willey & Sons, INC. 1941.

[23] N. Omori, H. Matsuo, S. Watanabe, M. Puschmann, Influence of carbon monoxide gas on silicon dioxide dry etching, Surf. Sci. 352-354 (1996) 988-992.

[24] M. Setareh, M. Farni, A. Maghari, A. Bogaerts, $CF_4$ decomposition in a low-pressure ICP: influence of applied power and $O_2$ content, J. Phys. D: Appl. Phys. 47 (2014) 355205 (15pp).

[25] L.M. Buchmann, F. Heinrich, P. Hoffmann, and J. Janes, Analysis of a $CF_4/O_2$ plasma using emission, laser-induced fluorescence, mass, and Langmuir spectroscopy, J. AppL Phys. 67 (1990) 3635-3640.

[26] Z. el Otell, V. Šamara, A. Zotovich, T. Hansen, J-F de Marneffe, M. R. Baklanov, Vacuum ultra-violet emission of $CF_4$ and $CF_3I$ containing plasmas and Their effect on low-k materials, J. Phys. D: Appl. Phys. 48 (2015) 395202 (9pp).

[27] M.S. Bak, W. Kim, M. A. Cappelli, On the quenching of excited electronic states of molecular nitrogen in nanosecond pulsed discharges in atmospheric pressure air, App. Phys. Lett. 98 (2011) 011502 (3pp).